\documentclass[sigconf]{acmart}

\usepackage{booktabs} % For formal tables

\usepackage{amsmath,amssymb,amsfonts}
\usepackage{graphicx}
\usepackage{textcomp}
\usepackage{xcolor}
\usepackage{color}
\usepackage{colortbl}
\usepackage{float}
\usepackage{graphicx}
\usepackage{subcaption}
\usepackage{algorithm}
\usepackage{algpseudocode}
\usepackage{mwe}

\usepackage{threeparttable}
\usepackage{multirow}

\newcommand{\ra}[1]{\renewcommand{\arraystretch}{#1}}

\makeatletter % https://tex.stackexchange.com/questions/33866/algorithm-tag-and-page-break
\newenvironment{breakablealgorithm}
  {% \begin{breakablealgorithm}
   \begin{center}
     \refstepcounter{algorithm}% New algorithm
     \hrule height.8pt depth0pt \kern2pt% \@fs@pre for \@fs@ruled
     \renewcommand{\caption}[2][\relax]{% Make a new \caption
       {\raggedright\textbf{\ALG@name~\thealgorithm} ##2\par}%
       \ifx\relax##1\relax % #1 is \relax
         \addcontentsline{loa}{algorithm}{\protect\numberline{\thealgorithm}##2}%
       \else % #1 is not \relax
         \addcontentsline{loa}{algorithm}{\protect\numberline{\thealgorithm}##1}%
       \fi
       \kern2pt\hrule\kern2pt
     }
  }{% \end{breakablealgorithm}
     \kern2pt\hrule\relax% \@fs@post for \@fs@ruled
   \end{center}
  }
\makeatother

\newcommand{\xoverbrace}[2][\vphantom{A}]{\overbrace{#1#2}}
\newcommand{\headersty}[1]{{\normalfont\normalsize\centering\scshape #1}}
\newcommand{\unaryminus}{\scalebox{0.75}[1.0]{\( - \)}}
\newcommand{\algrule}[1][.2pt]{\par\vskip.5\baselineskip\hrule height #1\par\vskip.5\baselineskip}

\copyrightyear{2019} 
\acmYear{2019} 
\setcopyright{acmlicensed}
\acmConference[CoNGA'19]{Conference for Next Generation Arithmetic 2019}{March 13--14, 2019}{Singapore, Singapore}
\acmBooktitle{Conference for Next Generation Arithmetic 2019 (CoNGA'19), March 13--14, 2019, Singapore, Singapore}
\acmPrice{15.00}
\acmDOI{10.1145/3316279.3316282}
\acmISBN{978-1-4503-7139-1/19/03}
\graphicspath{{./figures/}}

\begin{document}
\title{Performance-Efficiency Trade-off of Low-Precision Numerical Formats in Deep Neural Networks}
\renewcommand{\shorttitle}{Performance-Efficiency Trade-off of Low-Precision Numerical Formats in DNNs}
% \titlenote{Produces the permission block, and
% \subtitle{Extended Abstract}
% \subtitlenote{The full version of the author's guide is available as
%   \texttt{acmart.pdf} document}

% NOTE: change line 2170 of acmart.cls to change the authors in "ACM Reference Format" (this is due to us not having an author block per author). Normally "\author" is in place of our names.

\author{\texorpdfstring{{{Zachariah Carmichael}}\textsuperscript{\S}, {{Hamed~F.~Langroudi}}\textsuperscript{\S}, {{Char~Khazanov}}\textsuperscript{\S}, {{Jeffrey~Lillie}}\textsuperscript{\S},\\\vspace{-2.5mm}{{John~L. ~Gustafson}}\textsuperscript{\textdagger}, {{Dhireesha~Kudithipudi}}\textsuperscript{\S}}{}}
% \author{\texorpdfstring{{{Zachariah~Carmichael}}\textsuperscript{\S}}{}}
% \author{\texorpdfstring{{{Hamed~F.~Langroudi}}\textsuperscript{\S}}{}}
% \author{\texorpdfstring{{{Char~Khazanov}}\textsuperscript{\S}}{}}
% \author{\texorpdfstring{{{Jeffrey~Lillie}}\textsuperscript{\S}}{}}
% \author{\texorpdfstring{{{John~L.~Gustafson}}\textsuperscript{\textdagger}}{}}
% \author{\texorpdfstring{{{Dhireesha~Kudithipudi}}\textsuperscript{\S}}{}}

\affiliation{%
  \institution{\textsuperscript{\S} {{Neuromorphic~AI~Lab, Rochester~Institute~of ~Technology, ~Rochester, ~NY-14623}}}
  %\city{Rochester}
  %\state{NY}
  %\postcode{14623}
}
\affiliation{%
  \institution{\textsuperscript{\textdagger} {{National ~University ~of ~Singapore, ~Singapore-119077}}}
  %\streetaddress{21 Lower Kent Ridge Road}
  %\city{}
 %\state{Singapore}
  %\postcode{119077}
}

%\email{zjc2920@rit.edu}
%\author{Hamed F. Langroudi}
% \authornote{Dr.~Trovato insisted his name be first.}
%\email{sf3052@rit.edu}
%\email{rxk3067@rit.edu}
%\email{jsliee@rit.edu}
%\author{}
% \authornote{Dr.~Trovato insisted his name be first.}
%\affiliation{%
%  \institution{National University of Singapore}
%  \streetaddress{21 Lower Kent Ridge Road}
%  \city{}
 % \state{Singapore}
 % \postcode{119077}
%}
%\email{john.gustafson@nus.edu.sg}
%\email{dxkeec@rit.edu}

\renewcommand{\shortauthors}{Z. Carmichael \textit{et al.}}

\begin{abstract}  % TODO: shorten title
Deep neural networks (DNNs) have been demonstrated as effective prognostic models across various domains, \textit{e.g.} natural language processing, computer vision, and genomics. However, modern-day DNNs demand high compute and memory storage for executing any reasonably complex task. To optimize the inference time and alleviate the power consumption of these networks, DNN accelerators with low-precision representations of data and DNN parameters are being actively studied. An interesting research question is in how low-precision networks can be ported to edge-devices with similar performance as high-precision networks.
% In this work, we employ the tapered precision posit numerical system at $\leq$8-bit precision within a DNN accelerator, Deep Positron, for inference. To mitigate the impact of quantization, we utilize exact multiply-and-accumulate (EMAC) units for the fixed-point, floating point, and posit numerical formats and analyze the units for latency, power, and resource utilization on a Virtex-7 FPGA.
In this work, we employ the fixed-point, floating point, and posit numerical formats at $\leq$8-bit precision within a DNN accelerator, Deep Positron, with exact multiply-and-accumulate (EMAC) units for inference.
A unified analysis quantifies the trade-offs between overall network efficiency and performance across five classification tasks. Our results indicate that posits are a natural fit for DNN inference, outperforming at $\leq$8-bit precision, and can be realized with competitive resource requirements relative to those of floating point.
\end{abstract}




\maketitle

\begin{figure}[H]
    \centering
    \begin{subfigure}{0.47\linewidth}
        \vspace{1mm}
        \includegraphics[width=\textwidth]{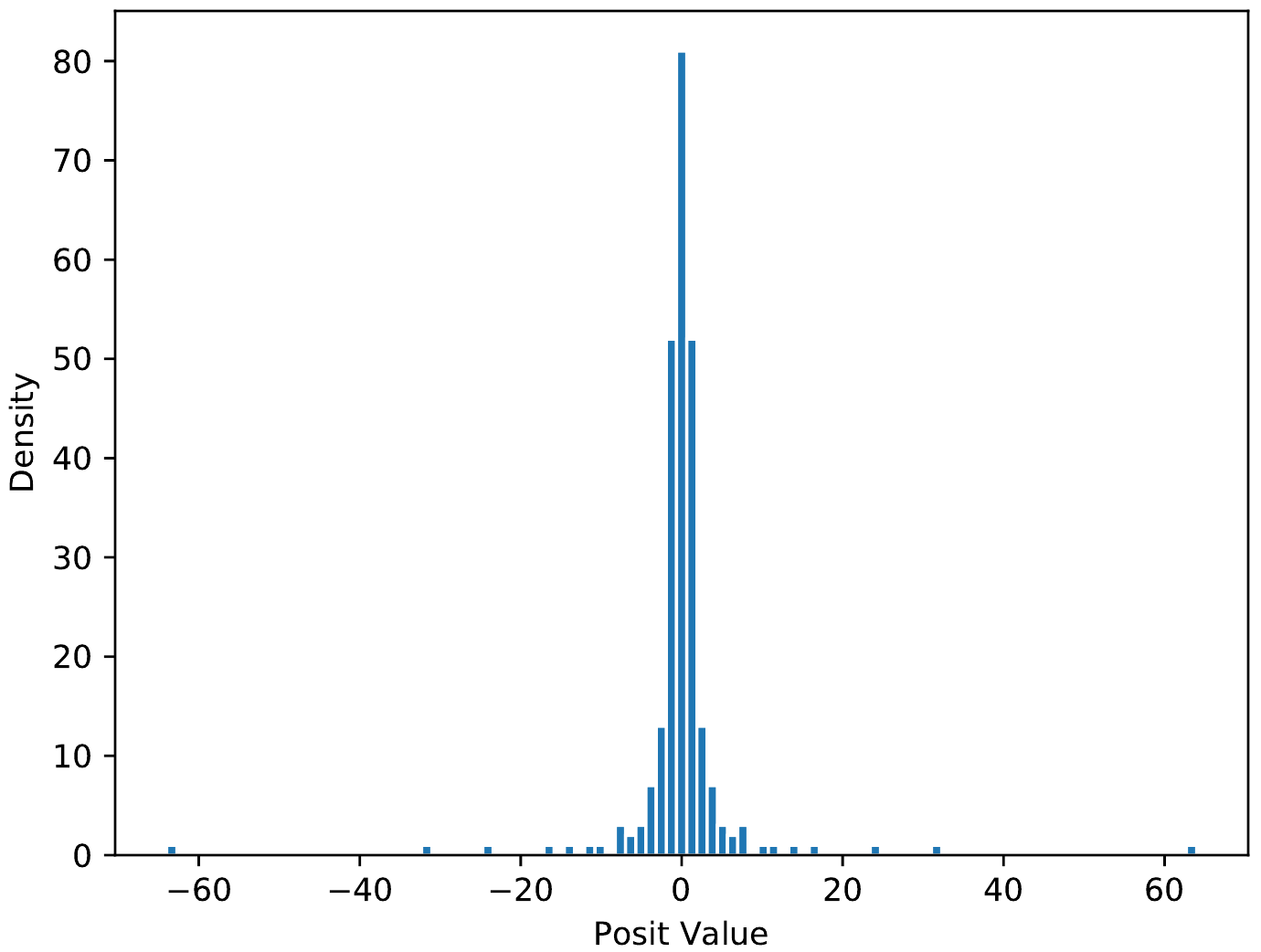}
    \end{subfigure}
    \begin{subfigure}{0.51\linewidth}
        \includegraphics[width=\textwidth]{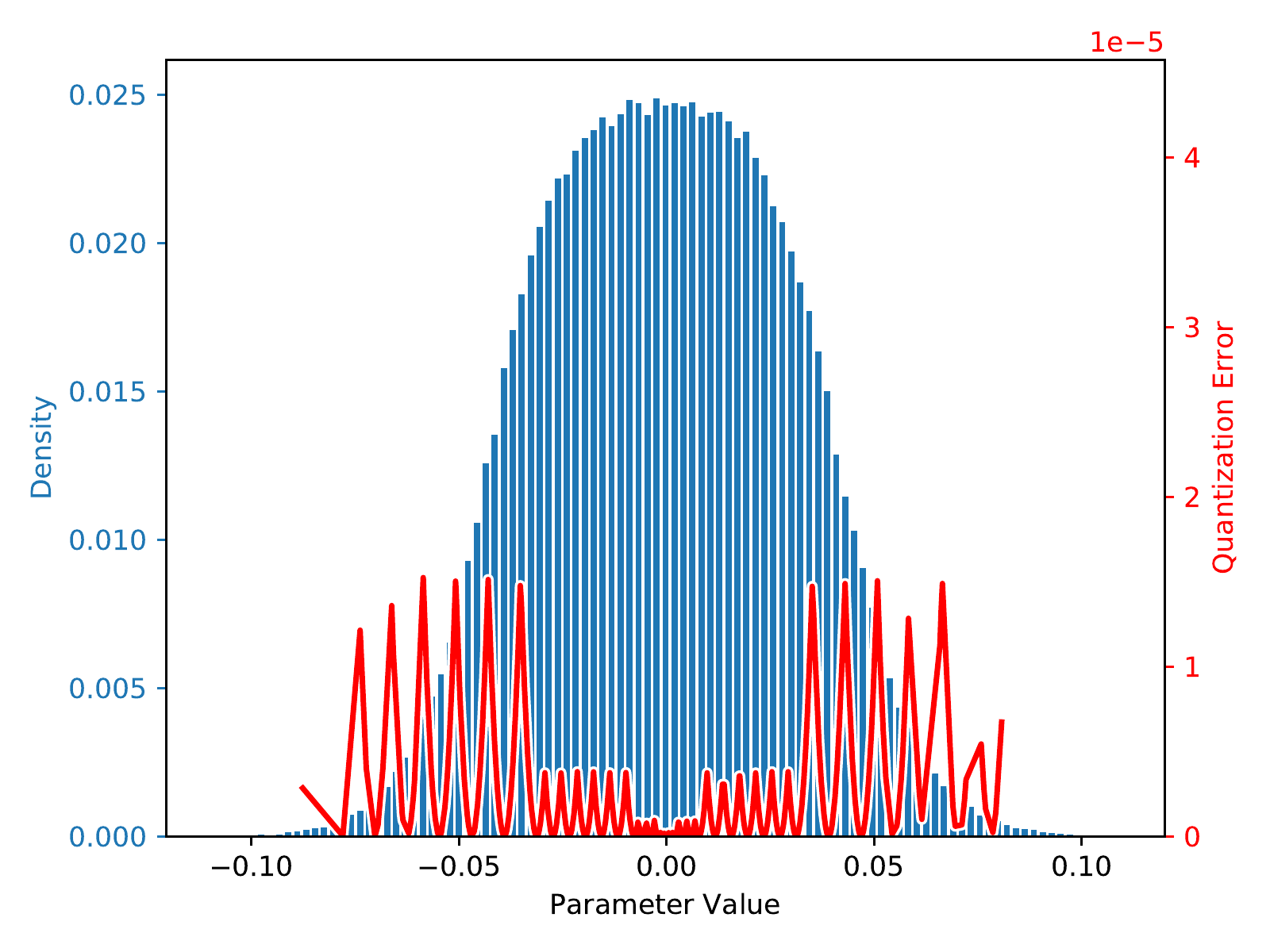}
    \end{subfigure}
    \caption{(a) 8-bit posit ($es=0$) value distribution; (b) Normalized ConvNet parameter distribution overlaid with quantization error (squared error). Both exhibit high density in the [-0.5,+0.5] range.}
    \label{fig:distrubution}
\end{figure}

\section{Introduction}

The deep neural network (DNN) is a popular learning paradigm that can generalize to tasks from disparate domains while achieving state-of-the-art performance. However, these networks are computationally heavyweight with regard to both compute and memory resources. For example, an outrageously large neural network with 32-bit floating point, such as an LSTM with a mixture of experts \cite{shazeer2017outrageously}, approximately requires 137 billion parameters. To manage the training and batch inference of these networks, hardware accelerators are employed, such as Google's Tensor, Processing Unit to decrease latency and increase throughput, embedded and/or reconfigurable devices to mitigate power bottlenecks, or targeted ASICs to optimize the overall performance. A predominant factor contributing to the computational cost is the large footprint of primitives, known as multiply-and-accumulate (MAC) operations, which perform weighted summations of the neuronal inputs. Techniques such as sparsity and low-precision representation \cite{DeepCompress2016,wu2018training,Microsoft2018,colangelo2018exploration} have been extensively studied to reduce the cost associated with MACs. For example, substituting 8-bit fixed-point for 32-bit fixed-point when performing inference on CIFAR-10 with AlexNet reduces the energy consumption 6$\times$ \cite{hashemi2017understanding}. These techniques become a necessity when deploying DNNs on end-devices, such as AI on the edge or IoT devices.

Of the methods used to mitigate these constraints, low-precision techniques have shown the most promise. For example, linear and nonlinear quantization have been able to match 32-bit floating point performance with 8-bit fixed-point and 8-bit floating point accelerators \cite{TPU2017,Minerva2016,Microsoft2018}. However, quantizing to an ultra-low bit precision, \textit{i.e.} $\leq$8-bits, can necessitate an increase in computational complexity. For example, a DNN has to be retrained or the number of hyperparameters significantly increased \cite{mishra2018wrpn} to maintain performance. A more lightweight solution is to perform DNN training and inference at a low-precision numerical format (fixed-point, floating point, or posit \cite{Ristretto}) instead of quantizing a trained network (\textit{e.g.} with 32-bit floating point). Previous studies have compared DNN inference with low-precision (\textit{e.g.} 8-bit) to high-precision floating point (\textit{e.g.} 32-bit) \cite{hashemi2017understanding}. However, these works compare numerical formats with disparate bit-widths and thereby do not fairly provide a comprehensive, holistic study of the network efficiency.

The recently proposed posit numerical format offers wider dynamic range, better accuracy, and improved closure over IEEE-754 floating point \cite{gustafson2017beating}. Fig.~\ref{fig:distrubution} shows intuitively that a natural posit distribution (\textit{e.g.} 8-bit posit, $es=0$) may be an optimal fit for representing DNN parameters (\textit{e.g.} of ConvNet). In this work, we investigate the effectiveness of ultra-low precision posits for DNN inference. The designs of several multiply-and-accumulate units for the posit, fixed-point, and floating point formats at low-precision are analyzed for resource utilization, latency, power consumption, and energy-delay-product. We carry out various classification tasks and compare the trade-offs between accuracy degradation and hardware efficacy. Our results indicate that posits outperform at ultra-low precision and can be realized at a similar cost to floating point in DNN accelerators.

\section{Related Work}
Since the late 1980s, low-precision fixed-point and floating point computation have been studied \cite{iwata1989artificial,hammerstrom1990vlsi}. In recent years, research attention has increased towards deep learning applications. Multiple groups have demonstrated that 16-bit fixed-point DNNs can perform inference with trivial degradation in performance \cite{Courbariaux14,bengio2013deep}. However, most of these works study DNN inference at varying bit-precision. There is a need for a more fair comparison between different number formats of corresponding bit-width paired with FPGA soft cores. For instance, Hashemi \textit{et al.} analyze 32-bit fixed-point and 32-bit floating point DNN inference on three DNN architectures (LeNet, ConvNet, and AlexNet) and show that fixed-point reduces the energy consumption by $\sim$12\% while suffering a mere 0--1\% accuracy drop~\cite{hashemi2017understanding}. Recently, Chung \textit{et al.} proposed a DNN accelerator (Brainwave) that increases inference throughput within a Stratix-10 FPGA by 3$\times$ by substituting 8-bit \textit{ms-fp8}, a novel spatial floating point format, in place of 8-bit fixed-point \cite{Microsoft2018}.

Several groups have previously studied the usage of the posit format in DNNs. Langroudi \textit{et al.} study the efficacy of posit representations of DNN parameters and activations \cite{Langroudi2018}. The work demonstrates that DNN inference using 7-bit posits endures $<$1\% accuracy degradation on ImageNet classification using AlexNet and that posits have a 30\% less ravenous memory footprint than fixed-point for multiple DNNs while maintaining a $<$1\% drop in accuracy. Cococcioni \textit{et al.} review the effectiveness of posits for autonomous driving functions \cite{cococcioni2018}. A discussion of a posit processing unit as an alternative to a floating point processing unit develops into an argument for posits as they exhibit a better trade-off between accuracy and implementation complexity. Most recently, J. Johnson proposed a log float format which couples posits with a logarithmic EMAC operation referred to as exact log-linear multiply-add (ELMA) \cite{johnson2018rethinking}. Use of the novel format within ResNet-50 achieves $<$1\% accuracy deterioration for ImageNet classification, and the ELMA shows much lower power consumption than the IEEE-754 floating point.

In this work, we demonstrate that posit arithmetic at ultra-low bit-width is an innate fit for DNN inference. The EMAC-equipped, parameterized Deep Positron architecture is mounted on an FPGA soft processor and compares assiduously the fixed-point, floating point, and posit formats at same bit-width.

\section{Background}

\subsection{Deep Neural Networks}
The DNN is a connectionist, predictive model used commonly for classification and regression. These networks learn a nonlinear input-to-output mapping in either a supervised, unsupervised, or semi-supervised manner. Before being able to perform inference, a DNN is trained to minimize a cost function and update parameters, called weights and biases, using backpropagation. Customarily, either 16-bit or 32-bit floating point arithmetic is used for DNN inference. However, 32-bit IEEE-754 floating point representation maintains a massive dynamic range of over 80 decades, which is beyond the range required for DNNs. Thus, this design of numerical distribution yields low information-per-bit based on Shannon maximum entropy \cite{shannon1948mathematical}. 16-bit floating point, often present in NVIDIA accelerators, unveils the format's limitations: nontrivial exception cases, underflow and overflow to $\pm$infinity or zero, and redundant NaN and zero representations. \emph{Posit arithmetic} offers an elegant solution to these limitations at generic bit-width.

\subsection{Posit Numerical Format}
The posit numerical format, a Type III unum, was proposed to improve upon the deficiencies of the IEEE-754 floating point format and to address complaints about Type I and II unums \cite{gustafson2017beating, tichy2016unums}. The posit format offers better dynamic range, accuracy, and program reproducibility than IEEE floating point. A posit number comprises $n$ bits and $es$ exponent bits, which controls the dynamic range. The primary divergence posit takes from floating point is the introduction of a signed, run-length encoded \emph{regime} bit-field. The longer this field is, a posit number has lower precision but larger magnitude, and vice versa for shorter run-lengths. Two posit bit-strings are reserved: $\tt{00}...\tt{0}$ for zero and $\tt{10}...\tt{0}$ for ``Not a Real,'' which can denote infinity, division by zero, \textit{etc.}. The following shows the interpretation of a binary posit bit-string.

\begin{equation*}\label{eq:posit}
  \underbrace{
    \xoverbrace{s}                      ^\textup{Sign}
    \xoverbrace{r~r~...~r~\bar{r}~}     ^\textup{Regime}
    \xoverbrace{e_1~e_2~e_3~...~e_{es}~}^\textup{Exponent, if any}
    \xoverbrace{f_1~f_2~f_3~...}        ^\textup{Mantissa, if any}
  }_{n~\textup{Bits}}
\end{equation*}

\noindent The numerical value a posit represents is then given by \eqref{eq:posit_value}

\begin{equation}\label{eq:posit_value}
  (-1) ^ s \times \left( 2 ^ {2 ^ {es}} \right) ^ k \times 2 ^ e \times 1.f
\end{equation}

\noindent where $k$ is the regime, $e$ is the unsigned exponent ($es > 0$), and $f$ is the value of the fraction bits. If a posit number is negative, the 2's complement is taken before decoding. We recommend reviewing \cite{gustafson2017beating} for a more thorough introduction and intuition to the posit format.

\section{Methodology}

We build off of \cite{carmichael2019positron}, using the proposed Deep Positron architecture. The framework is parameterized by bit-width, numerical type, and DNN hyperparameters, so networks of arbitrary width and depth can be constructed for the fixed-point, floating point, and posit formats. The following sections further describe the EMAC operation and detail the EMAC algorithms for each numerical format.

\subsection{Exact Multiply-and-Accumulate (EMAC)}

The multiply-and-accumulate (MAC) operation is ubiquitous within DNNs -- each neuron computes a weighted sum of its inputs. In most implementations, this operation is usually inexact, meaning rounding or truncation results in accumulation of error. The EMAC mitigates this issue by implementing a variant of the Kulisch accumulator \cite{kulisch2013computer} and delaying error until every product of each layer has been accumulated. This minimization of local error becomes substantial at low-precision. In each EMAC module, a wide register accumulates fixed-point values and rounds in a deferred stage. For $k$ multiplications, the accumulator width is computed using \eqref{eq:kulisch_width}
\begin{equation}\label{eq:kulisch_width}
	w_a = \lceil \log_2(k) \rceil + 2 \times \left\lceil \log_2 \left(\frac{max}{min} \right) \right\rceil + 2
\end{equation}
\noindent where $max$ and $min$ are the maximum and minimum value magnitudes for a given numerical system, respectively. Each EMAC is pipelined into three stages: multiplication, accumulation, and rounding. A fourth stage, implementing the trivial activation function, $\textup{ReLU}(x)=\textup{max}(x,0)$, is present for hidden layer neurons. For further introduction to EMACs and the exact dot product, we recommend reviewing \cite{kulisch2013computer, carmichael2019positron}.

\begin{figure}[H]
  \centering
  \includegraphics[width=\linewidth]{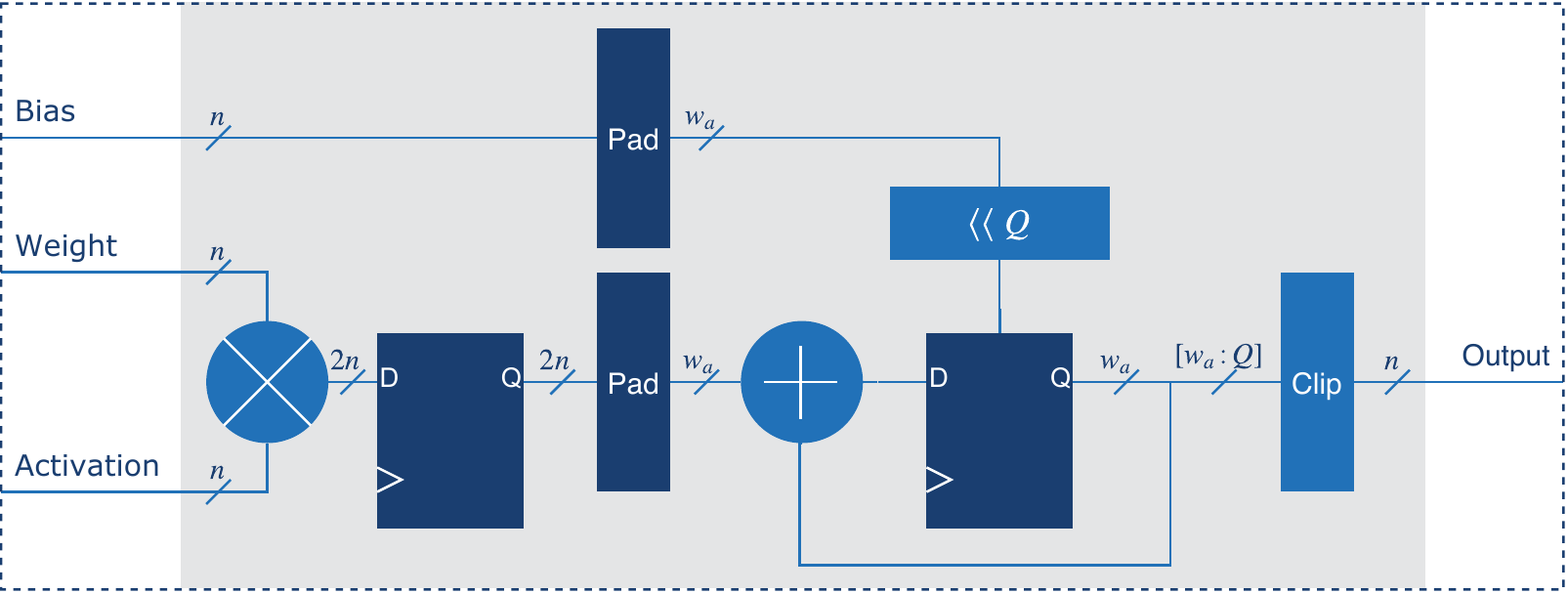}
  \caption{A parameterized ($n$ total bits, $Q$ fractional bits) FPGA soft core design of the fixed-point exact multiply-and-accumulate operation \cite{carmichael2019positron}.}
  \label{fig:fixed_mac}
\end{figure}

\subsection{Fixed-Point EMAC}

We parameterize the fixed-point EMAC as $n$, the bit-width, and $Q$, the number of fractional bits, where $n > Q$. Fig. \ref{fig:fixed_mac} shows the block diagram design of the EMAC with signal bit-widths indicated. The functionality of the unit is described by Algorithm \ref{alg:fixed}. The general characteristics of a fixed-point number are given by the following.

\begin{align*}
max &= 2 ^ {-Q} \times (2 ^ {n-1} - 1)\\
min &= 2 ^ {-Q}
\end{align*}

\begin{algorithm}
%   \small
  \caption{Fixed-point EMAC operation.}
  \label{alg:fixed}
  \begin{algorithmic}[1]
    \setlength{\thinmuskip}{2mu}
    \setlength{\medmuskip}{3mu plus 1.5mu minus 3mu}
    \setlength{\thickmuskip}{3.5mu plus 3.5mu}
    \Procedure{FixedEMAC}{{\tt weight}, {\tt activation}}
    \algrule \hspace{-1.75mm}\textbf{Multiplication}
    \State ${\tt prod} \leftarrow {\tt weight} \times {\tt activation}$
    \algrule \hspace{-1.75mm}\textbf{Accumulate}
    \State ${\tt sum} \leftarrow {\tt prod} + {\tt sum}$
    \algrule \hspace{-1.75mm}\textbf{Rounding and Clipping}
    \If {$({\sim}{\tt sum}[{\tt MSB}]) \& |{\tt sum}[{\tt MSB}\unaryminus{}1:n+Q]$}
      \State ${\tt sum} \leftarrow 2 ^ {n - 1} - 1$\Comment{Set to max pos. value}
    \ElsIf {${\tt sum}[{\tt MSB}] \& ({\sim}(\&{\tt sum}[{\tt MSB}\unaryminus{}1:n+Q\unaryminus{}1]))$}
      \State ${\tt sum} \leftarrow -(2 ^ {n - 1})$\Comment{Set to min neg. value}
    \Else
      \State ${\tt sum} \leftarrow \textup{round}({\tt sum})$
    \EndIf
    \algrule \hspace{-1.75mm}\textbf{Normalize}
    \State ${\tt result} \leftarrow ({\tt sum} \gg Q)[n-1:0]$
    \EndProcedure
  \end{algorithmic}
\end{algorithm}

% ===============================================================
\subsection{Floating Point EMAC}

\begin{figure*}
  \centering
  \includegraphics[width=.88\linewidth]{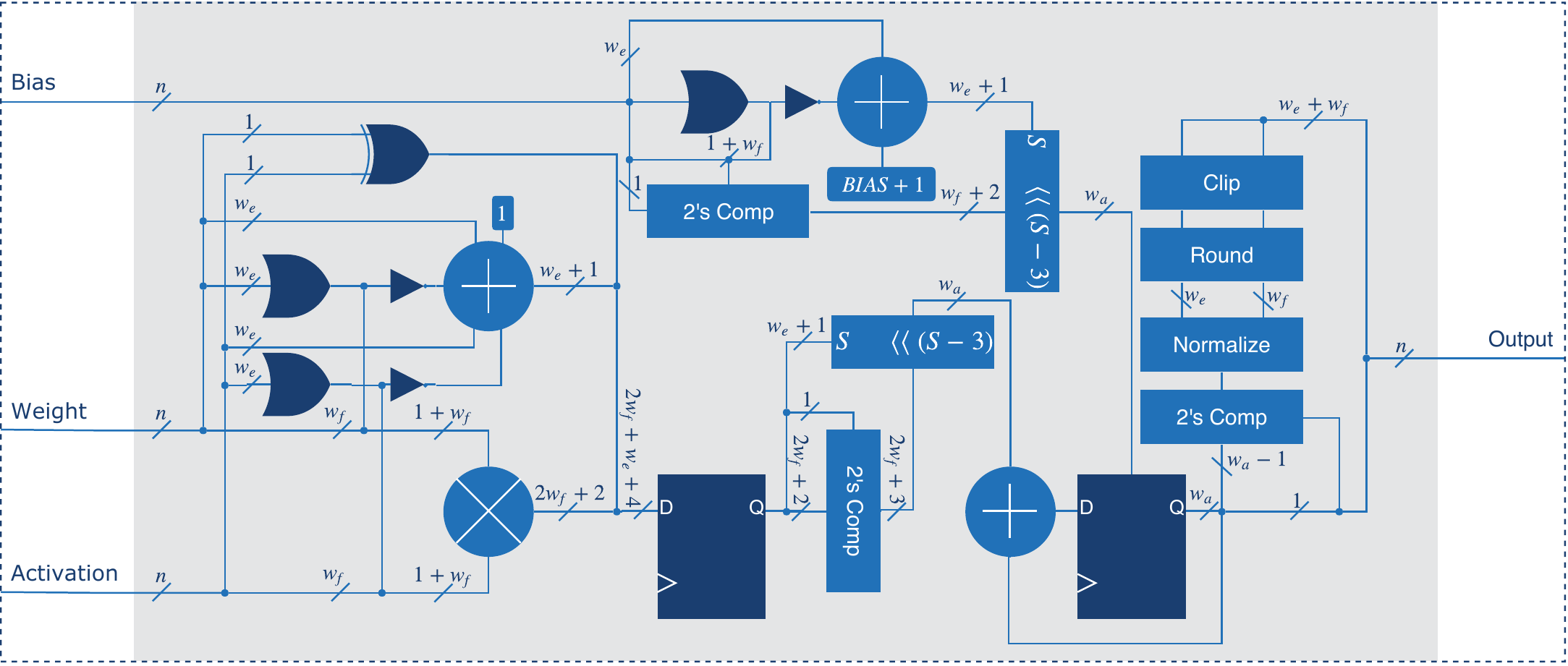}
  \caption{A parameterized ($n$ total bits, $w_e$ exponent bits, $w_f$ fractional bits) FPGA soft core design of the floating-point exact multiply-and-accumulate operation \cite{carmichael2019positron}.}
  \label{fig:float_mac}
\end{figure*}

The floating point EMAC is parameterized by $w_e$, the number of exponent bits, and $w_f$, the number of fractional bits. As all inputs and intermediate values in Deep Positron are real-valued, we do not consider ``Not a Number'' (NaN) or ``$\pm$ Infinity'' in this implementation. Fig. \ref{fig:float_mac} shows the floating point EMAC block diagram with labeled bit-widths of signals. A leading-zeros-detector (LZD) is used in converting from fixed-point back to floating point. The EMAC functionality is expressed in Algorithm \ref{alg:float}, and the relevant characteristics of the floating point format are computed as follows.

\begin{align*}
bias &= 2 ^ {w_e - 1} - 1\\
exp_{max} &= 2 ^ {w_e} - 2\\
max &= 2 ^ {exp_{max} - bias} \times (2 - 2 ^ {-w_f})\\
min &= 2 ^ {1 - bias} \times 2 ^ {-w_f}
\end{align*}

\begin{breakablealgorithm}%[H]
%   \small
  \caption{Floating point EMAC operation.}
  \label{alg:float}
  \begin{algorithmic}[1]
  \begingroup
    \setlength{\thinmuskip}{2mu}
    \setlength{\medmuskip}{3mu plus 1.5mu minus 3mu}
    \setlength{\thickmuskip}{3.5mu plus 3.5mu}
    \Procedure{FloatEMAC}{${\tt{s_{wgt}}},{\tt{m_{wgt}}},{\tt{e_{wgt}}},{\tt{s_{act}}},{\tt{m_{act}}},{\tt{e_{act}}}$}%\Comment{Data extraction of $\tt{in}$}
        \algrule \hspace{-1.75mm}\textbf{Subnormal Detection}
        \State ${\tt zero_{wgt}} \leftarrow |{\tt e_{wgt}}$
        \State ${\tt zero_{act}} \leftarrow |{\tt e_{act}}$
        \State ${\tt ms_{wgt}} \leftarrow {\sim}{\tt zero_{wgt}} \& {\tt m_{wgt}}$\Comment{Add hidden bit to mantissa}
        \State ${\tt ms_{act}} \leftarrow {\sim}{\tt zero_{act}} \& {\tt m_{act}}$\Comment{Add hidden bit to mantissa}
        \algrule \hspace{-1.75mm}\textbf{Multiplication}
        \State ${\tt{s_p}} \leftarrow {\tt{s_{wgt}}} \oplus {\tt{s_{act}}}$
        \State ${\tt{m_p}} \leftarrow {\tt{ms_{wgt}}} \times {\tt{ms_{act}}}$
        \State ${\tt{e_p}} \leftarrow {\tt{e_{wgt}}} + {\tt{e_{act}}} + {\tt zero_{wgt}} + {\tt zero_{act}} + 1$
        \algrule \hspace{-1.75mm}\textbf{Conversion to Fixed-Point}
        \State ${\tt m_{fx}} \leftarrow {\tt s_p}~?~\unaryminus{}{\tt m_p}:{\tt m_p}$
        \State ${\tt m_{fx}} \leftarrow {\tt m_{fx}} \ll ({\tt e_p} - 3)$\Comment{Min exponent is 3}
        \algrule \hspace{-1.75mm}\textbf{Accumulate}
        \State ${\tt sum} \leftarrow {\tt m_{fx}} + {\tt sum}$
        \algrule \hspace{-1.75mm}\textbf{Convert Back to Floating Point\footnotemark}%\protect
        \State ${\tt s_r} \leftarrow {\tt sum}[{\tt MSB}]$
        \State ${\tt mag} \leftarrow {\tt s_r}~?~\unaryminus{}{\tt sum}:{\tt sum}$
        \State ${\tt ovf} \leftarrow {\sim}|{\tt sum}[{\tt MSB}:exp_{max}\unaryminus 3]$
        \State ${\tt zc} \leftarrow \textup{LZD}({\tt mag})$
        \State ${\tt m_r},{\tt guard},{\tt sticky} \leftarrow {\tt mag}[{\tt MSB}\unaryminus{}{\tt zc}:{\tt MSB}\unaryminus{}{\tt zc}\unaryminus{}w_f\unaryminus{}2]$
        \State ${\tt e_r} \leftarrow exp_{max}\unaryminus{}{\tt zc}$\Comment{Zeros count is the biased exponent}
        \State ${\tt lsb} \leftarrow {\tt m_r}[0]$
        \State ${\tt rc} \leftarrow {\tt guard}~\&~({\tt lsb}|{\tt sticky})$\Comment{Round check}
        \State ${\tt m_r} \leftarrow {\tt m_r} + {\tt rc}$
        \State ${\tt result} \leftarrow \{{\tt s_r}, {\tt e_r}, {\tt m_r}\}$
    \EndProcedure
  \endgroup
  \end{algorithmic}
\end{breakablealgorithm}

\footnotetext{Note that during the conversion back to floating point overflow handling is omitted for simplicity.}

\subsection{Posit EMAC}

\begin{figure*}
  \centering
  \includegraphics[width=\linewidth]{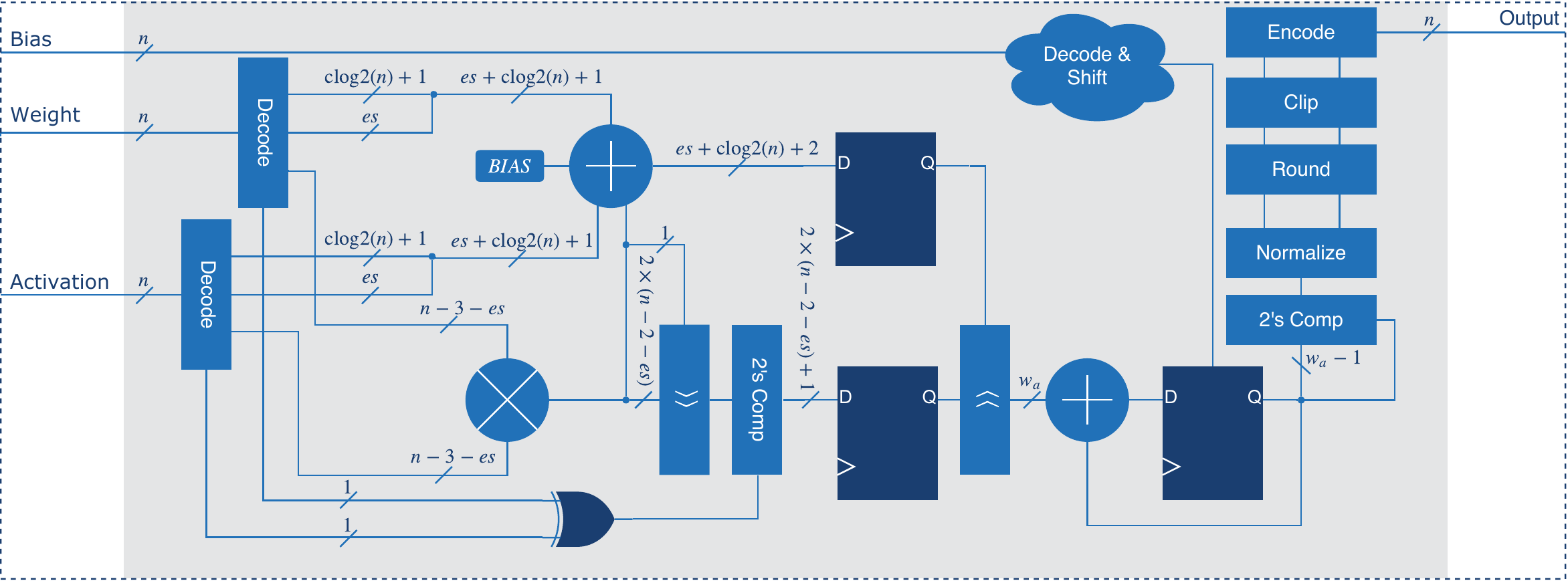}
  \caption{A parameterized ($n$ total bits, $es$ exponent bits) FPGA soft core design of the posit exact multiply-and-accumulate operation \cite{carmichael2019positron}.}
  \label{fig:posit_mac}
%   \vspace{-5mm}
\end{figure*}

The posit EMAC, shown in Fig. \ref{fig:posit_mac}, is parameterized by $n$, the bit-width, and $es$, the number of exponential bits. In this implementation, we do not consider ``Not a Real'' as all DNN parameters and data are real-valued and posits do not overflow to infinity. Algorithm \ref{alg:posit_data_extract} describes the data extraction process for each EMAC input, which is more involved per the dynamic length regime. The EMAC employs this process as outlined by Algorithm \ref{alg:posit_edp}. The relevant attributes of a given posit format are calculated using the following,
\begin{align*}
useed  &= 2^{2^{es}} \\
max &= useed^{n-2} \\
min &= useed^{-n+2}
\end{align*}
where $useed$ can be thought of as the scale factor base, as shown in \eqref{eq:posit_value}.

\vspace{7mm}

\begin{algorithm}[t]
  \caption{Posit data extraction of $n$-bit input with $es$ exponent bits}\label{alg:posit_data_extract}
  \begin{algorithmic}[1]
  	\begingroup
    %   \small
      \setlength{\thinmuskip}{2mu}
      \setlength{\medmuskip}{3mu plus 1.5mu minus 3mu}
      \setlength{\thickmuskip}{3.5mu plus 3.5mu}
      \Procedure{Decode}{$\tt{in}$}\Comment{Data extraction of $\tt{in}$}
        \State ${\tt{nzero}} \gets |{\tt{in}}$\Comment{`1' if $\tt{in}$ is nonzero}
        \State ${\tt{sign}} \gets {\tt{in}}[n\unaryminus1]$\Comment{Extract sign}
        \State ${\tt{twos}} \gets (\{n\unaryminus1\{{\tt{sign}}\}\}\oplus {\tt{in}}[n\unaryminus2:0])+{\tt{sign}}$\Comment{2's Comp.}
        \State ${\tt{rc}} \gets {\tt{twos}}[n\unaryminus2]$\Comment{Regime check}
        \State ${\tt{inv}} \gets \{n\unaryminus1\{{\tt{rc}}\}\} \oplus {\tt{twos}}$\Comment{Invert 2's}
        \State ${\tt{zc}} \gets \textup{LZD}({\tt{inv}})$\Comment{Count leading zeros}
        \State ${\tt{tmp}} \gets {\tt{twos}}[n\unaryminus4:0] \ll ({\tt{zc}}-1)$\Comment{Shift out regime}
        \State ${\tt{frac}} \gets \{{\tt{nzero}}, {\tt{tmp}}[n\unaryminus es \unaryminus 4:0]\}$\Comment{Extract fraction}
        \State ${\tt{exp}} \gets {\tt{tmp}}[n\unaryminus 4:n\unaryminus es \unaryminus 3]$\Comment{Extract exponent}
        \State ${\tt{reg}} \gets {\tt{rc}}~?~{\tt{zc}}\unaryminus1:\unaryminus {\tt{zc}}$\Comment{Select regime}
        \State \textbf{return} $\tt{sign}$, $\tt{reg}$, $\tt{exp}$, $\tt{frac}$
      \EndProcedure
    \endgroup
  \end{algorithmic}
\end{algorithm}

% Change to breakablealgorithm instead of algorithm to break between pages
\begin{breakablealgorithm}%[H]
  \caption{Posit EMAC operation for $n$-bit inputs each with $es$ exponent bits}\label{alg:posit_edp}
  \begin{algorithmic}[1]
  	\begingroup
    %   \small
      \setlength{\thinmuskip}{2mu}
      \setlength{\medmuskip}{3mu plus 1.5mu minus 3mu}
      \setlength{\thickmuskip}{3.5mu plus 3.5mu}
      \Procedure{PositEMAC}{$\tt{weight,activation}$}
        \State ${\tt{sign_w}}, {\tt{reg_w}}, {\tt{exp_w}}, {\tt{frac_w}} \gets \text{\headersty{Decode}}({\tt{weight}})$
        \State ${\tt{sign_a}}, {\tt{reg_a}}, {\tt{exp_a}}, {\tt{frac_a}} \gets \text{\headersty{Decode}}({\tt{activation}})$
        \State ${\tt{sf_w}} \gets \{{\tt{reg_w}}, {\tt{exp_w}}\}$\Comment{Gather scale factors}
        \State ${\tt{sf_a}} \gets \{{\tt{reg_a}}, {\tt{exp_a}}\}$
        \algrule \hspace{-1.75mm}\textbf{Multiplication}
        \State ${\tt{sign_{mult}}} \gets {\tt{sign_w}} \oplus {\tt{sign_a}}$
        \State ${\tt{frac_{mult}}} \gets {\tt{frac_w}} \times {\tt{frac_a}}$
        \State ${\tt{ovf_{mult}}} \gets {\tt{frac_{mult}}}[{\tt{MSB}}]$\Comment{Adjust for overflow}
        \State ${\tt{normfrac_{mult}}} \gets {\tt{frac_{mult}}} \gg {\tt{ovf_{mult}}}$
        \State ${\tt{sf_{mult}}} \gets {\tt{sf_{w}}} + {\tt{sf_{a}}} + {\tt{ovf_{mult}}}$
        \algrule \hspace{-1.75mm}\textbf{Accumulation}
        \State ${\tt{fracs_{mult}}} \gets {\tt{sign_{mult}}}~?~{\unaryminus\tt{frac_{mult}}}:{\tt{frac_{mult}}}$
        \State ${\tt{sf_{biased}}} \gets {\tt{sf_{mult}}} + bias$\Comment{Bias the scale factor}
        \State ${\tt{fracs_{fixed}}} \gets {\tt{fracs_{mult}}} \ll {\tt{sf_{biased}}}$\Comment{Shift to fixed}
        \State ${\tt{sum_{quire}}} \gets {\tt{fracs_{fixed}}} + {\tt{sum_{quire}}}$\Comment{Accumulate}
        \algrule \hspace{-1.75mm}\textbf{Fraction \& SF Extraction}
        \State ${\tt{sign_{quire}}} \gets {\tt{sum_{quire}}}[{\tt{MSB}}]$
        \State ${\tt{mag_{quire}}} \gets {\tt{sign_{quire}}}~?~{\unaryminus\tt{sum_{quire}}}:{\tt{sum_{quire}}}$
        \State ${\tt{zc}} \gets \textup{LZD}({\tt{mag_{quire}}})$
        \State ${\tt{frac_{quire}}} \gets {\tt{mag_{quire}}}[2{\times}(n\unaryminus 2\unaryminus es)\unaryminus 1{+}{\tt{zc}}:{\tt{zc}}]$
        \State ${\tt{sf_{quire}}} \gets {\tt{zc}} \unaryminus bias$
        %         \State ${\tt{ovf_{quire}}} \gets |{\tt{mag_{quire}}}[{\tt{MSB}}:{\tt{MSB}}\unaryminus 1]$
        \algrule \hspace{-1.75mm}\textbf{Convergent Rounding \& Encoding}
        \State ${\tt{nzero}} \gets |{\tt{frac_{quire}}}$
        \State ${\tt{sign_{sf}}} \gets {\tt{sf_{quire}}}[{\tt{MSB}}]$
    	\State ${\tt{exp}} \gets {\tt{sf_{quire}}}[es\unaryminus 1:0]$\Comment{Unpack scale factor}
        \State ${\tt{reg_{tmp}}} \gets {\tt{sf_{quire}}}[{\tt{MSB}}\unaryminus1:es]$
        \State ${\tt{reg}} \gets {\tt{sign_{sf}}}~?~\unaryminus{\tt{reg_{tmp}}}:{\tt{reg_{tmp}}}$
        \State ${\tt{ovf_{reg}}} \gets {\tt{reg}}[{\tt{MSB}}]$\Comment{Check for overflow}
        \State ${\tt{reg_f}} \gets {\tt{ovf_{reg}}}~?~\{\{\lceil\log_2(n)\rceil\unaryminus 2 \{{\tt{1}}\}\}), {\tt{0}}\} : {\tt{reg}}$
        \State ${\tt{exp_f}} \gets ({\tt{ovf_{reg}}}|{\sim}{\tt{nzero}}|{(\tt{\&{\tt{reg_f}}}}))~?~\{es\{{\tt{0}}\}\}:{\tt{exp}}$
        \State ${\tt{tmp1}} \gets \{{\tt{nzero}}, {\tt{0}}, {\tt{exp_f}}, {\tt{frac_{quire}}}[{\tt{MSB}}\unaryminus 1:0],\{n\unaryminus 1 \{{\tt{0}}\}\}\}$
        \State ${\tt{tmp2}} \gets \{{\tt{0}}, {\tt{nzero}}, {\tt{exp_f}}, {\tt{frac_{quire}}}[{\tt{MSB}}\unaryminus 1:0],\{n\unaryminus 1 \{{\tt{0}}\}\}\}$
        \State ${\tt{ovf_{regf}}} \gets \&{\tt{reg_f}}$
        \If {${\tt{ovf_{regf}}}$}
        	\State ${\tt{shift_{neg}}} \gets {\tt{reg_{f}}} - 2$
			\State ${\tt{shift_{pos}}} \gets {\tt{reg_{f}}} - 1$
		\Else
			\State ${\tt{shift_{neg}}} \gets {\tt{reg_{f}}} - 1$
			\State ${\tt{shift_{pos}}} \gets {\tt{reg_{f}}}$
		\EndIf
        \State ${\tt{tmp}} \gets {\tt{sign_{sf}}}~?~{\tt{tmp2}} \gg {\tt{shift_{neg}}} : {\tt{tmp1}} \gg {\tt{shift_{pos}}}$
        \State ${\tt{lsb}}, {\tt{guard}} \gets {\tt{tmp}}[{\tt{MSB}}\unaryminus (n\unaryminus 2):{\tt{MSB}}\unaryminus (n\unaryminus 1)]$
        \State ${\tt{round}} \gets {\sim}({\tt{ovf_{reg}}}|{\tt{ovf_{regf}}})~? $ \par
        $~~~~~~~~~~~~~~~~~~(~{\tt{guard}}~\&~({\tt{lsb}}~|~(|{\tt{tmp}}[{\tt{MSB}}\unaryminus n : 0]))~) : {\tt{0}}$
        \State ${\tt{result_{tmp}}} \gets {\tt{tmp}}[{\tt{MSB}} : {\tt{MSB}}\unaryminus n{+}1] {+} {\tt{round}}$
		\State ${\tt{result}} \gets {\tt{sign_{quire}}}~?~\unaryminus{\tt{result_{tmp}}} : {\tt{result_{tmp}}}$
        \State \textbf{return} ${\tt{result}}$
      \EndProcedure
    \endgroup
  \end{algorithmic}
\end{breakablealgorithm}

\begin{figure*}
\centering
\begin{subfigure}{.075\linewidth}
  \centering
  \caption{}
\end{subfigure}%
\begin{subfigure}{.39\linewidth}
  \centering
  \includegraphics[width=\linewidth]{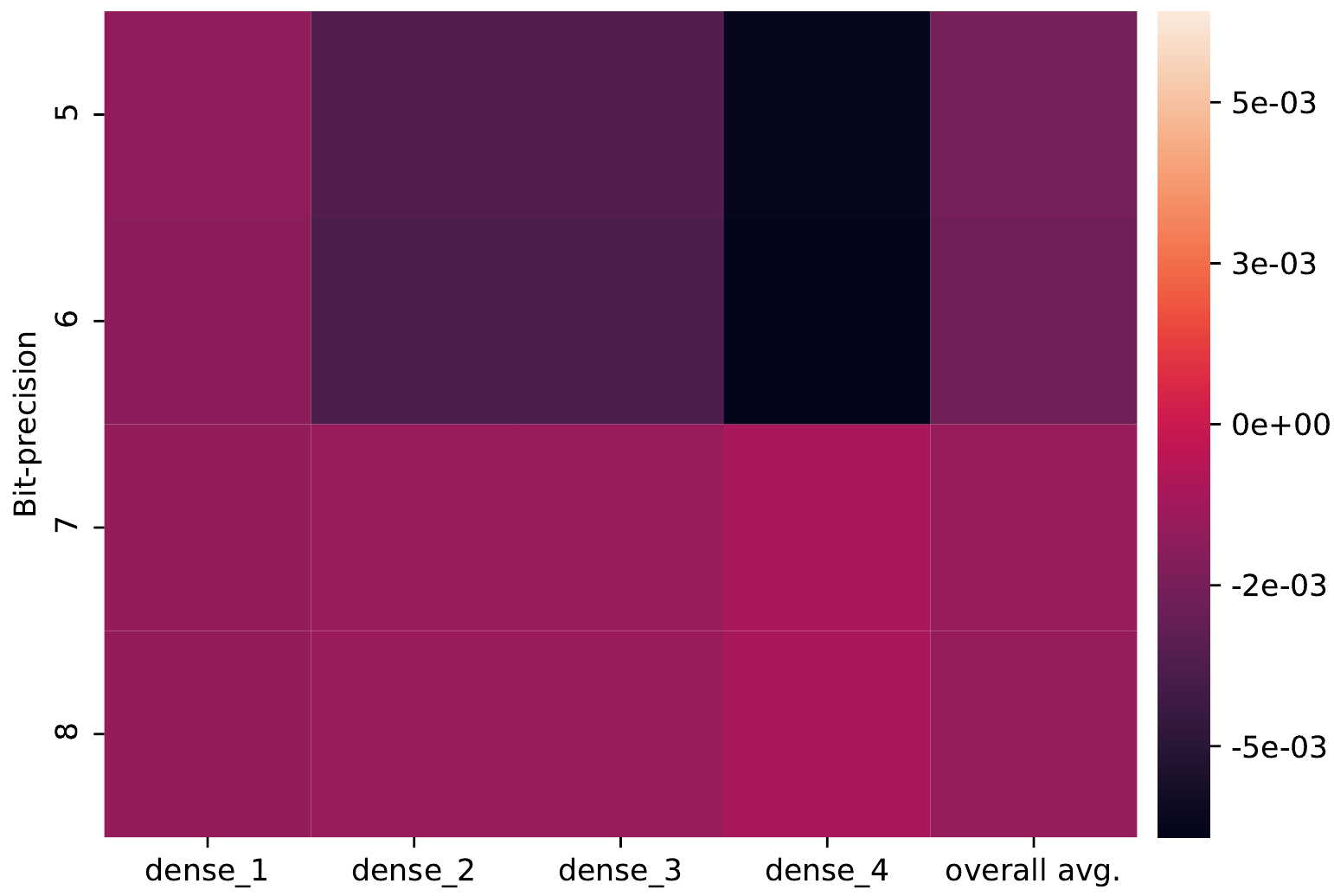}
\end{subfigure}%
\begin{subfigure}{.075\linewidth}
  \centering
  \caption{}
\end{subfigure}%
\begin{subfigure}{.39\linewidth}
  \centering
  \includegraphics[width=\linewidth]{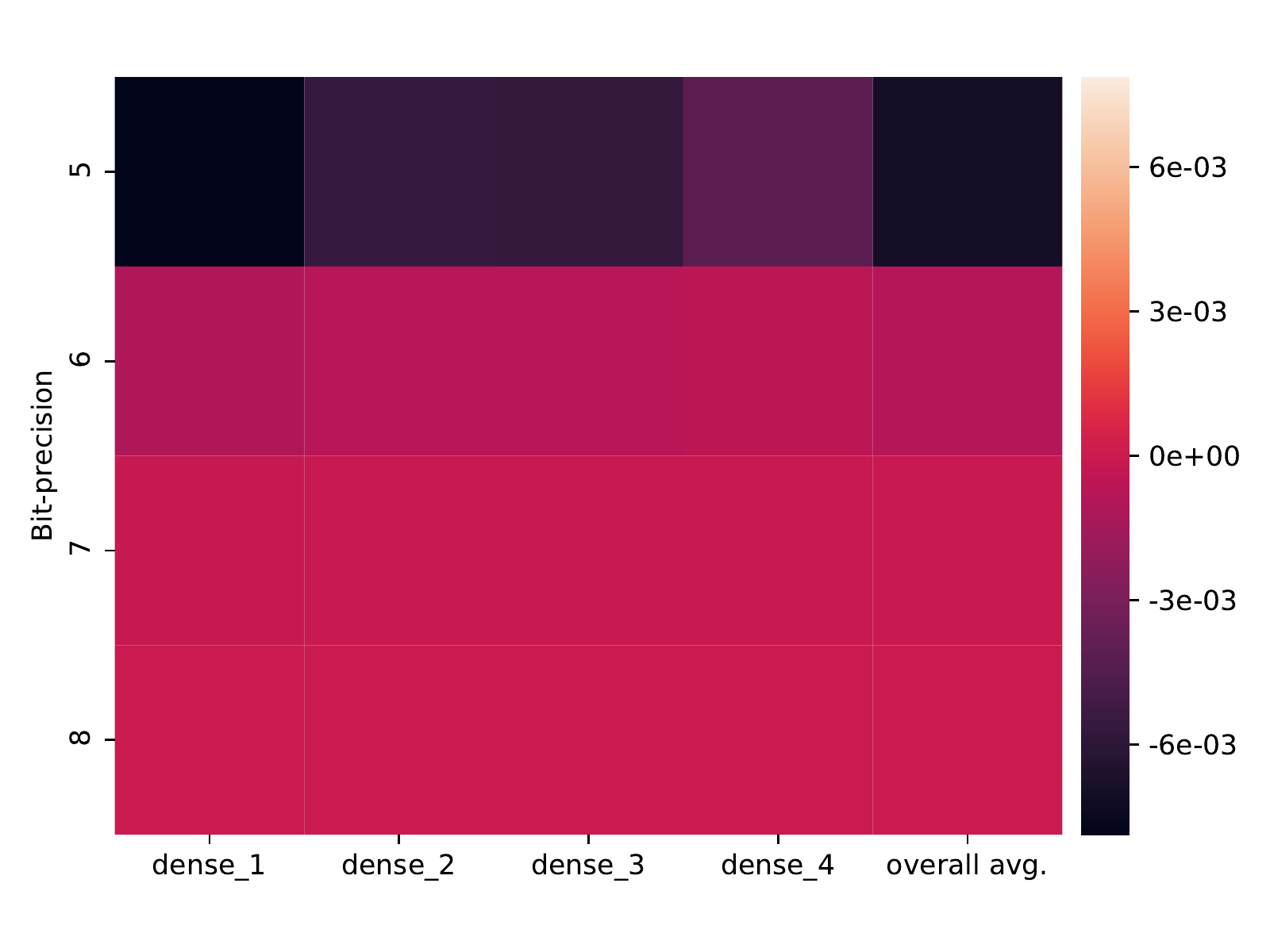}
\end{subfigure}\vskip\baselineskip\vspace{-3mm}
\begin{subfigure}{.075\linewidth}
  \centering
  \caption{}
\end{subfigure}%
\begin{subfigure}{.39\linewidth}
  \centering
  \includegraphics[width=\linewidth]{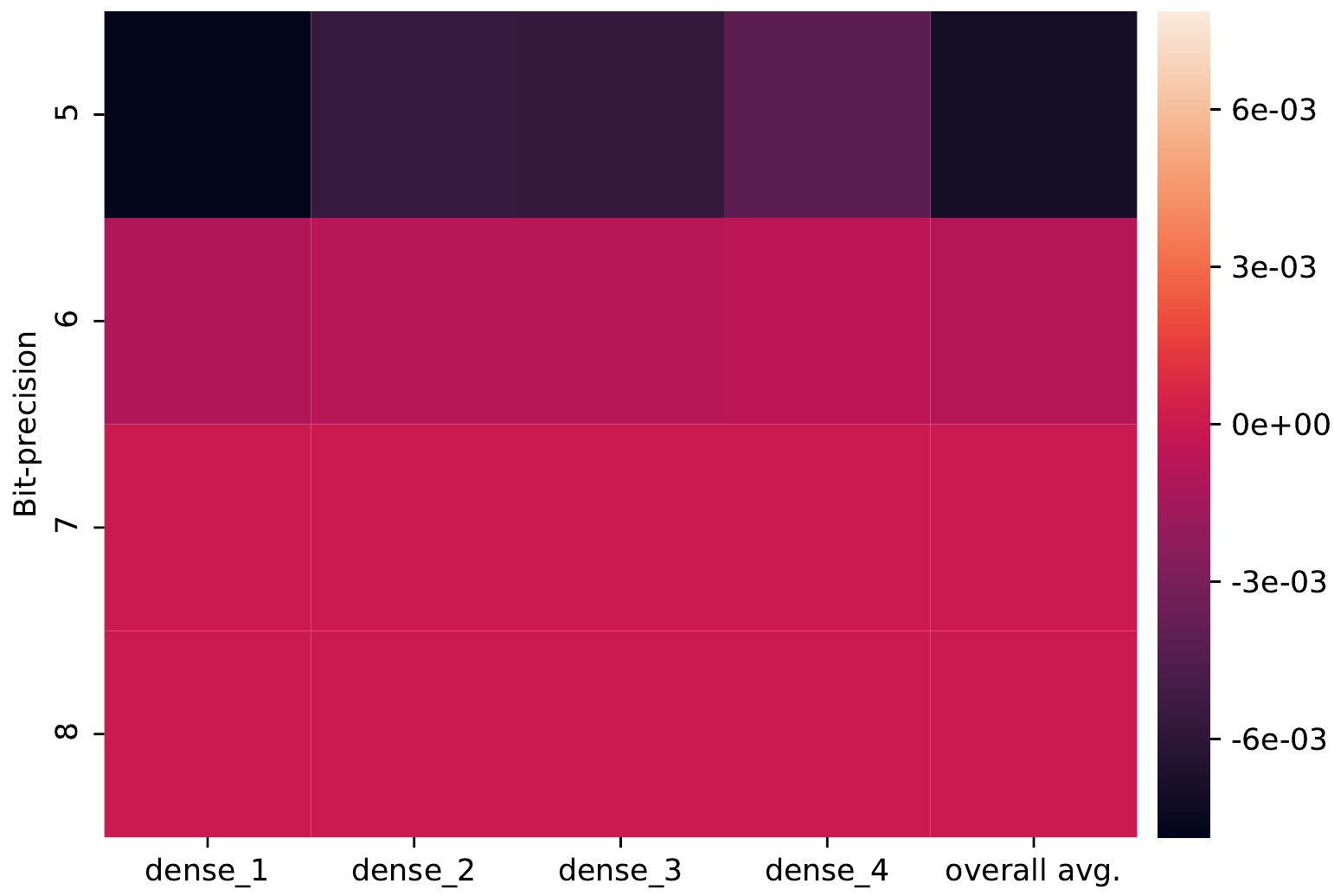}
\end{subfigure}%
\begin{subfigure}{.075\linewidth}
  \centering
  \caption{}
\end{subfigure}%
\begin{subfigure}{.39\linewidth}
  \centering
  \includegraphics[width=\linewidth]{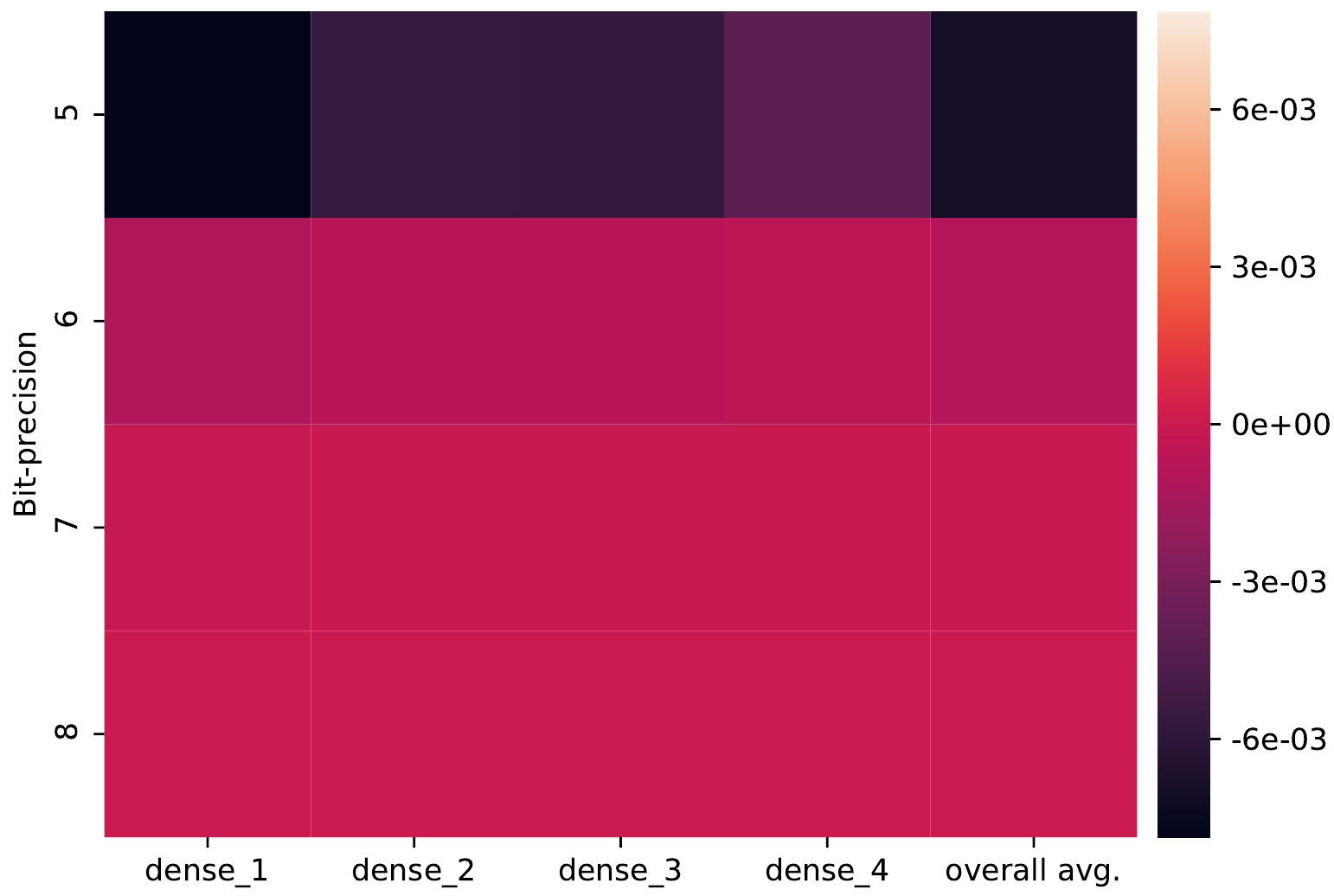}
\end{subfigure}
\caption{Layer-wise\protect\footnotemark~~quantization error (MSE) heatmaps compare the fitness of $[5,8]$-bit numerical formats with best performance of sweeping $es$, $w_e$ and $Q$ parameters for representing 32-bit floating point DNN parameters. The last column of each heatmap indicates the average quantization error among all parameters in a DNN. (a) $\textup{MSE}_{posit}-\textup{MSE}_{fixed}$ for the MNIST task; (b) $\textup{MSE}_{posit}-\textup{MSE}_{float}$ for the MNIST task; (c) $\textup{MSE}_{posit}-\textup{MSE}_{fixed}$ for the Fashion MNIST task; (d) $\textup{MSE}_{posit}-\textup{MSE}_{float}$ for the Fashion MNIST task.}
\label{fig:mse}
\end{figure*}

\section{Experimental Results}

% \subsection{EMAC Analysis and Comparison}
In all experiments, we synthesize the EMACs onto a Virtex-7 FPGA (xc7vx485t-2ffg1761c) using Vivado 2017.2. and expand upon the results from \cite{carmichael2019positron}. With regard to energy and latency, the posit EMAC is competitive with the floating point EMAC. While using more resources for the same bit-precision, posits offer a wider dynamic range at fewer bits while maintaining a faster maximum operational frequency. Moreover, the energy-delay-product of the floating point and posit EMACs are comparable. The fixed-point EMAC, obviously, is uncontested with its resource utilization and latency; its lack of an exponential parameter results in a far more slender accumulation register. However, fixed-point offers poor dynamic range compared with the other formats at the same bit-precision.

\begin{table}[b]
\caption{Deep Positron inference accuracy (Acc.) on the five low-dimensional datasets with 8-bit EMACs. The best results arise when posit has $es \in \{0, 1, 2\}$, floating point has $w_e \in \{3, 4\}$, and fixed-point has $Q \in \{4, 5\}$.}
\centering
\ra{1.2}
\resizebox{\linewidth}{!}{%
\begin{threeparttable}
\begin{tabular}{@{}ccccccc@{}}
\toprule
    %\cmidrule{4-7}
    &&&  & Floating & Fixed- & 32-bit\\[-1ex]
    &&& \multirow{-2}{*}{Posit} & Point & Point & Float\\ \cmidrule{4-7}
    \multirow{-3}{*}{Dataset} & \multirow{-3}{*}{\shortstack{Inference\\Size}} && \multicolumn{1}{c}{Acc. ($es$)} & \multicolumn{1}{c}{Acc. ($w_e$)} & \multicolumn{1}{c}{Acc. ($Q$)} & \multicolumn{1}{c}{Acc.} \\
    \hline
    WI Breast Cancer~\cite{WBC} & 190 && \textbf{85.9}\% (2) & 77.4\% (4) & 57.8\% (5) & 90.1\% \\
    % \hline
     Iris~\cite{IRIS} & 50 && \textbf{98.0}\% (1) & 96.0\% (3) & 92.0\% (4) & 98.0\%\\
    % \hline
     Mushroom~\cite{Mashrom}& 2,708 && \textbf{96.4}\% (1) & \textbf{96.4}\% (4) & 95.9\% (5) & 96.8\%\\
    % \hline
     MNIST \cite{lecun1998mnist} & 10,000 && \textbf{98.5}\% (1) & 98.4\% (4) & 98.3\% (5) & 98.5\% \\
    % \hline
     Fashion MNIST \cite{xiao2017fashion} & 10,000 && \textbf{89.6}\% (1) & \textbf{89.6}\% (4) & 89.2\% (4) & 89.5\%\\
    \bottomrule
\end{tabular}
% }
\begin{tablenotes}
\small
\item[2]{The term ``dense'' is synonymous with a fully-connected feedforward layer in a DNN.}
\end{tablenotes}
\end{threeparttable}
}
\label{table:1-1}
\end{table}

The quantization error of a tensor $X$ is computed as the mean-squared-error as shown in \eqref{eq:mse}.

\begin{equation}\label{eq:mse}
    \textup{MSE}(X,X_{quant}) = \frac{1}{n}\sum^n_i\left(X-X_{quant}\right)^2
\end{equation}

\noindent Fig. \ref{fig:mse} shows a layer-wise heatmap of quantization error between formats for the MNIST and Fashion MNIST classification tasks. It is clear that posits suffer the least consequences from quantization, which is especially noticeable at $\leq$5-bit precision.

We evaluate the inference accuracy of several feedforward three- or four-layer neural networks, instantiated on the Deep Positron accelerator, on five datasets. The baseline results are taken from networks trained and evaluated using standard IEEE-754 floating point at 32-bit precision.
The inputs and weights of the trained networks are quantized from the 32-bit floating point format to the desired numerical format (either $[5,8]$-bit posit, $[5,8]$-bit floating point, or $[5,8]$-bit fixed-point) via round-to-nearest with ties to even. The best performance is selected among $[5,8]$-bit formats with a sweep of the $es$, $w_e$, and $Q$ parameters for the posit, floating point, and fixed-point formats, respectively. Across all tasks, posit either outperforms or matches the performance of fixed-point and floating point, as shown in Table \ref{table:1-1}. In some cases, an 8-bit posit matches the performance of the 32-bit floating point baseline. An interesting result is that both posit and floating point at 8-bit precision improve upon the baseline performance for the Fashion MNIST task.

\begin{figure*}%[H]
    \centering
    \includegraphics[width=.9\linewidth]{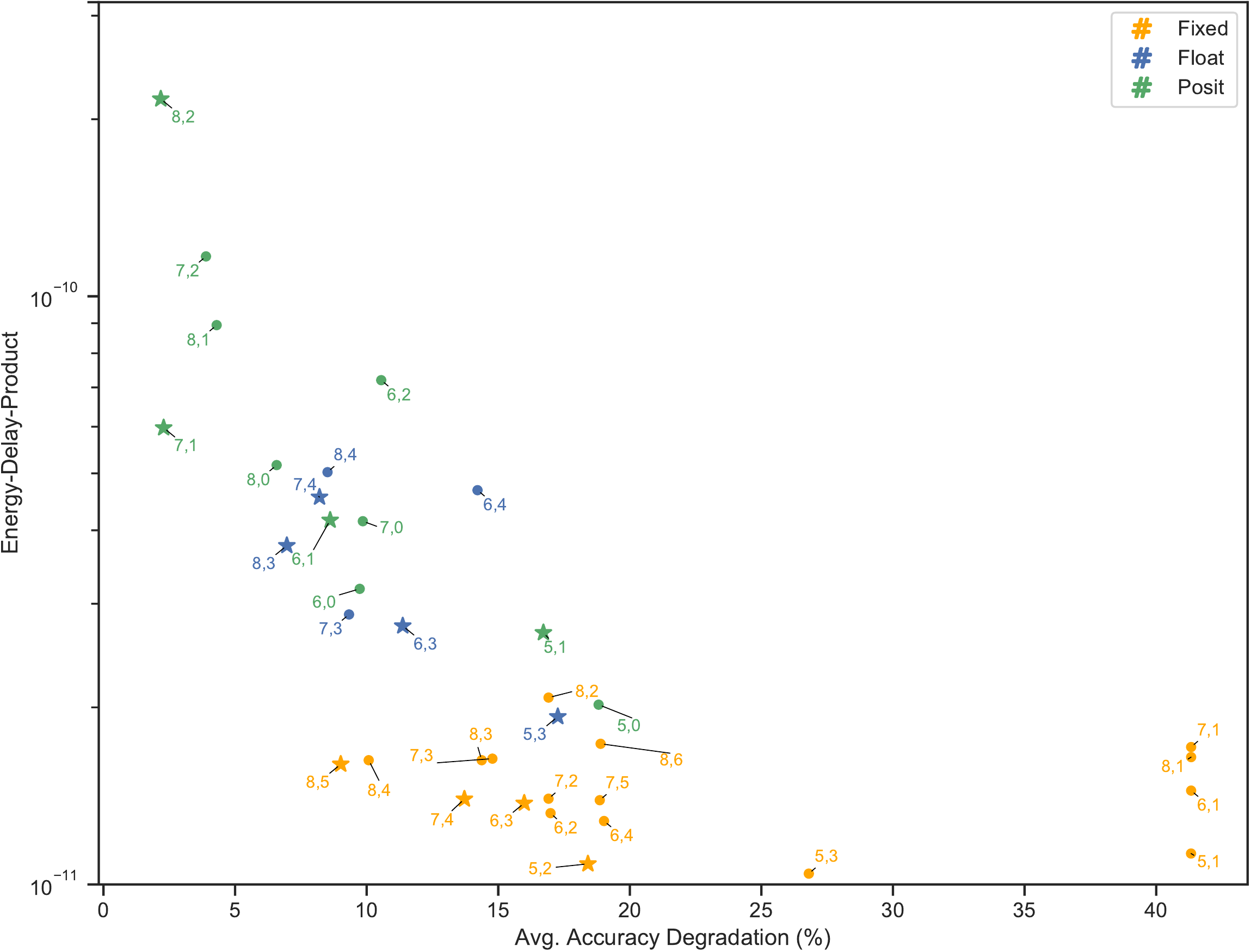}
    \caption{The average accuracy degradation from 32-bit floating point across the five classification tasks vs. the energy-delay-product of the respective EMAC. A star ($\star$) denotes the lowest accuracy degradation for a numerical format and bit-width.}
    \label{fig:acc_deg}
\end{figure*}

\begin{figure*}
\centering
\begin{subfigure}{.5\linewidth}
  \centering
  \includegraphics[width=\linewidth]{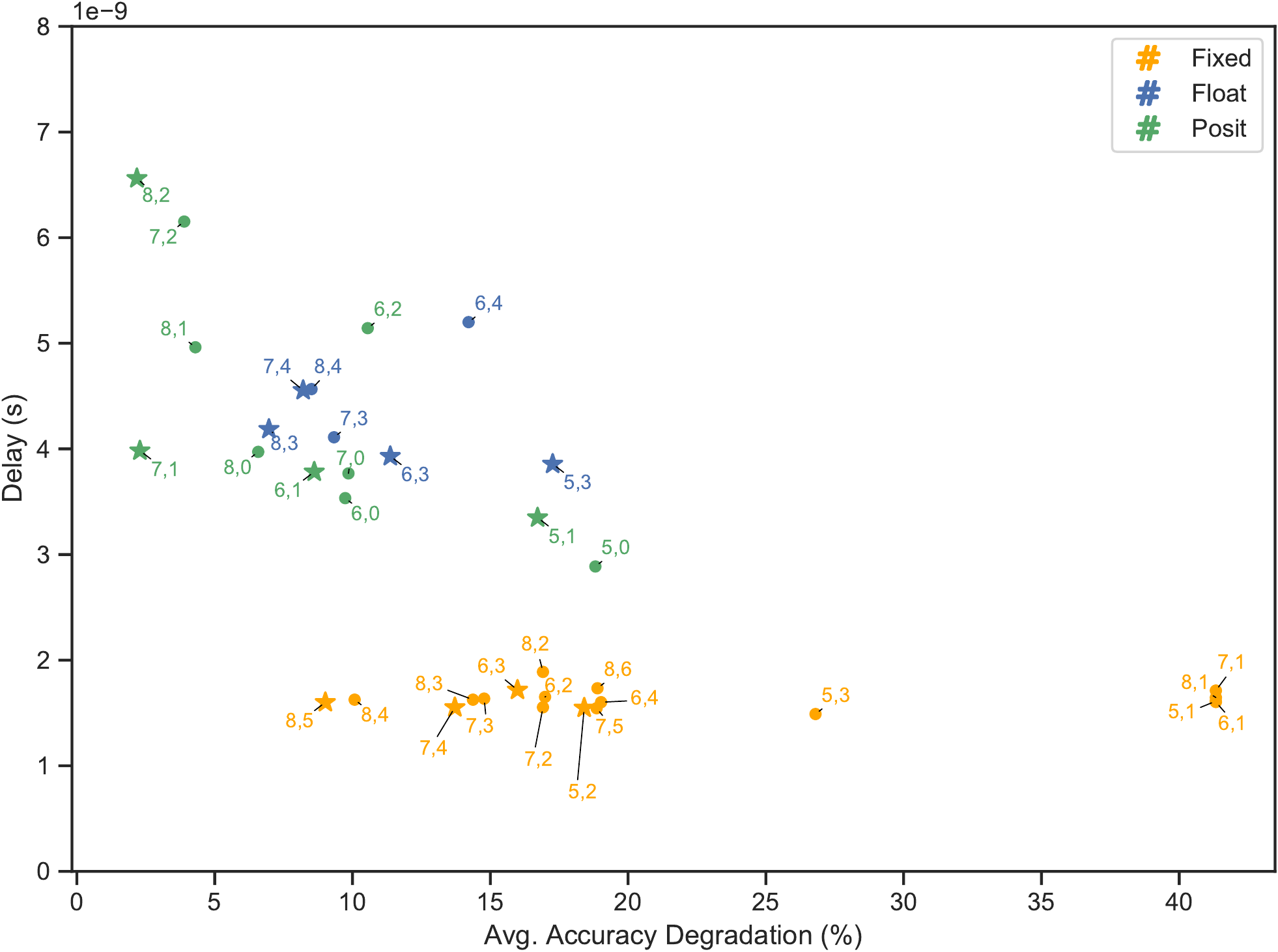}
\end{subfigure}%
\begin{subfigure}{.5\linewidth}
  \centering
  \includegraphics[width=\linewidth]{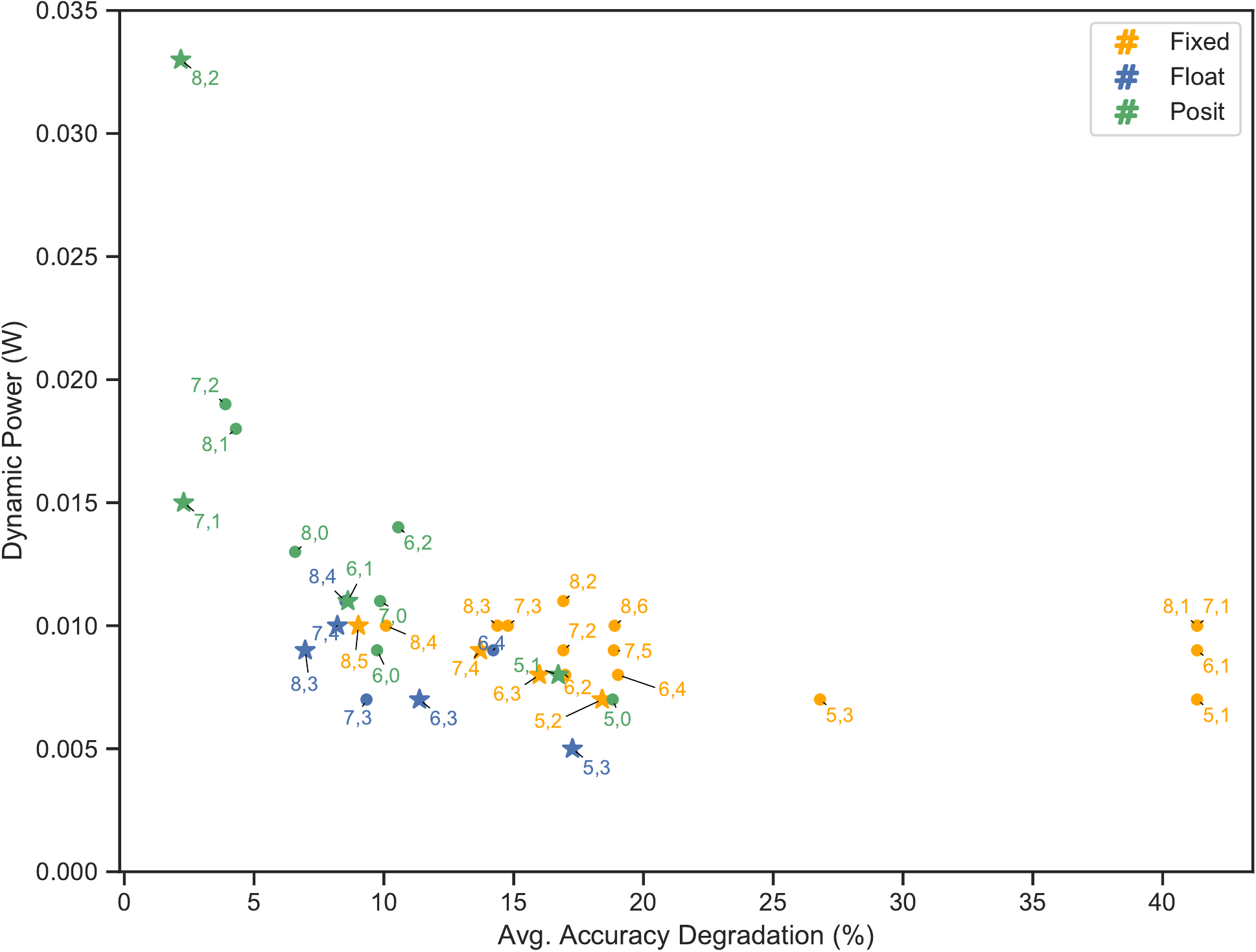}
\end{subfigure}%
\caption{The average accuracy degradation from 32-bit floating point across the five classification tasks vs. the delay (left) and the dynamic power (right) of the respective EMAC. A star ($\star$) denotes the lowest accuracy degradation for a numerical format and bit-width.}
\label{fig:acc_deg2}
\end{figure*}

We compare energy, delay, and the energy-delay-product against the average Deep Positron performance across all formats with $[5,8]$ bit-precision. Figs. \ref{fig:acc_deg} and \ref{fig:acc_deg2} depict the average accuracy degradation across the five classification tasks against these metrics for each bit-width. Posit consistently outperforms at a slight cost in power. Fixed-point maintains the lowest delay across all bit-widths, as expected, but offers the worst performance. While the floating point EMAC generally uses less power than the posit EMAC, the posit EMAC enjoys lower latencies across all bit-widths whilst maintaining lower accuracy degradation.

\begin{table*}
   \caption{Comparison of posit arithmetic hardware implementations.}
   \label{tab:regime}
   \ra{1.3}
   \centering
   \resizebox{\linewidth}{!}{%
   \begin{tabular}{@{}cccccccc@{}}
     \toprule
     Design & \cite{KumarDate}& \cite{Parameterized} & \cite{podobas} & \cite{IBM} & \cite{lehoczky2018high} & \cite{johnson2018rethinking} & This Work \\
     \midrule
     \multirow{2}{*}{Device} & \multirow{2}{*}{Virtex-6 FPGA/ASIC} & \multirow{2}{*}{Zynq-7000 SoC/ASIC} & Stratix V GX & Virtex7 VX690 \& Ultrascale & \multirow{2}{*}{Artix-7 FPGA} & \multirow{2}{*}{ASIC} & Virtex-7 (xc7vx485t-2ffg1761c)  \\[-1ex]
     &&& 5SGXA7 FPGA & Plus VU3P FPGAs &&& FPGA \\
    %  \hline
     Task & - & FIR Filter & - & - & - & Image Classification & Image Classification \\
    %  \hline
     \multirow{2}{*}{Dataset} & \multirow{2}{*}{-} & \multirow{2}{*}{-} & \multirow{2}{*}{-} & \multirow{2}{*}{-} & \multirow{2}{*}{-} & \multirow{2}{*}{ImageNet} & WI Breast Cancer, Iris, Mush- \\[-1ex]
     &&&&&&& room, MNIST, Fashion MNIST \\
    %  \hline
     Bit-precision & All & All & All & 32 & All & All, emphasized on 8 & All, emphasized on $[5,8]$ \\ 
    %  \hline
     Operations & Mul,Add/Sub & Mul,Add/Sub & Mul,Add/Sub & Quire & Quire & Quire & Quire \\
    %  \hline
     Programming Language & Verilog & Verilog & C++ /OpenCL & Verilog & C\# & OpenCl & VHDL \\
    %  \hline
     Technology Node & 40 nm / 90 nm & 28 nm / 90 nm & 28 nm & 28 nm / 20 nm & 28 nm & 28 nm & 28 nm \\
     \bottomrule
   \end{tabular}
   }
   \label{tab:OthersWork}
 \end{table*}

\subsection{Exploiting the Posit \texorpdfstring{$es$}{es} Parameter}

Experimental results in this paper are evaluated by exploiting the performance of posit numerical formats with $es \in \{0,1,2\}$ across five data sets. As is shown in Fig. 6, the energy-delay-product of the posit EMAC is dependent upon the $es$ parameter. For instance, the energy-delay-product of the posit EMAC with $es=0$, on average, is 3{$\times$} and 1.4{$\times$} less than the energy-delay-product of the posit EMAC with $es=2$ and $es=1$, respectively. On the other hand, the average performance of DNN inference with $es=1$ for the posit EMAC among the five datasets and $[5,7]$ bit-precision is 2\% and 4\% percent better than with $es=2$ and $es=0$, respectively. Thus, Deep Positron equipped with the posit ($es=1$) EMAC has a better trade-off between energy-delay-product and accuracy for $[5,7]$ bits. For 8-bit, the results suggest that $es=1$ is a better fit for energy-efficient applications and $es=2$ for accuracy-dependent applications.

% ================

\subsection{Comparison with Other Posit Hardware Implementations}

A summary of previous studies which design posit arithmetic hardware is shown in Table \ref{tab:OthersWork}. Several groups implement posit basic arithmetic algorithms, such as addition, subtraction, multiplication, and exact-dot-product (Quire) on FPGA for various applications \cite{KumarDate,KumarISCAS,Parameterized,IBM,podobas,lehoczky2018high,johnson2018rethinking}.
Kumar \textit{et al.} provided a hardware generator for posit addition, subtraction, and multiplication and showed reduced latency and area consumption of 32-bit posit addition with $es=3$ over IEEE-754 floating point addition \cite{KumarDate,KumarISCAS}.
However, the comparison is between two different FPGA platforms which diminishes the merit of this comparison. They also ignore several characteristic demands for posit arithmetic, such as round-to-nearest with ties to even or unbiased rounding. To better realize the advantages of posit arithmetic over IEEE-754 floating point with complete posit arithmetic features, Chaurasiya \textit{et al.} proposed a parameterized posit arithmetic hardware generator \cite{Parameterized}. They emphasized that resource utilization and energy of the posit arithmetic unit is comparable with IEEE-754 float when the same number of bits are considered for both formats. However, the area consumption of the posit hardware is less than IEEE-745 float at similar precision and dynamic range. To simplify and expedite hardware design, as well as improve the usability of posits on heterogeneous platforms, researchers in \cite{lehoczky2018high} and \cite{podobas} use high-level languages, such as C\# and OpenCL, to generate posit arithmetic hardware for FPGAs.

Most of the previous works do not support the exact-dot-product operation and do not design specialized posit arithmetic for deep learning applications as we presented in this paper. In \cite{carmichael2019positron}, a parameterized FPGA-mounted DNN accelerator is constructed which employs exact-dot-product algorithms for the posit, fixed-point, and floating point formats. The paper shows strong preliminary results that posits are a natural fit for low-precision inference. Proceeding this work, J. Johnson proposed an exact log-linear multiply-add arithmetic algorithm for deep learning applications using a posit multiplier in the log domain and a Kulisch adder \cite{johnson2018rethinking}. The results indicate better performance of 8-bit posit multiply-add over 8-bit fixed-point multiply-add with similar accuracy for the ResNet-50 neural network and ImageNet dataset. However, the paper targets an ASIC platform and convolutional neural network at 8-bit precision whereas we study an FPGA implementation and fully-connected neural network at $[5,8]$ bit-precision.

\section{Conclusions}

We demonstrate that the recent posit numerical system has a high affinity for deep neural network inference at $\leq$8-bit precision. The proposed posit hardware is shown to be competitive with the floating point counterpart in terms of resource utilization and energy-delay-product. Moreover, the posit EMAC offers a superior maximum operating frequency over that of floating point. With regard to performance degradation, direct quantization to ultra-low precision favors posits heavily, surpassing fixed-point vastly. Moreover, the performance of floating point is either matched or surpassed consistently by posits across multiple datasets. The success of prospective new classes of learning algorithms will be coordinately contingent on the underlying hardware.

\bibliographystyle{ACM-Reference-Format}
\bibliography{sample-sigconf}

\end{document}